\documentclass[12pt]{article}

\usepackage[dvips]{graphicx}

\newcommand{\nn}{\nonumber}
\newcommand{\be}{\begin{equation}}
\newcommand{\ee}{\end{equation}}
\newcommand{\ba}{\begin{eqnarray}}
\newcommand{\ea}{\end{eqnarray}}
\newcommand{\req}[1]{(\ref{#1})}
\def\={\,=\,}
\newcommand{\ci}[1]{\cite{#1}}
\def\tw{\textwidth}

\def\mev{~{\rm MeV}}
\def\gev{~{\rm GeV}}

\def\ale{\alpha_{\rm elm}}
\def\als{\alpha_{\rm s}}
\def\et1{\widetilde{\eta}_1}
\def\et8{\widetilde{\eta}_8}

\def\etp{\eta^\prime}

\def\ub{\bar{u}}
\def\db{\bar{d}}
\def\sb{\bar{s}}

\def\BbB{B\overline{B}}

\def\ov{\overline}

\begin{document} 
\thispagestyle{empty}
\begin{flushright}
WU B 13-16 \\
October, 24 2013\\[20mm]
\end{flushright}

\begin{center}

{\Large\bf Probing moments of baryon-antibaryon generalized parton distributions  
at BELLE and FAIR}\\
\vskip 10mm

P.\ Kroll$^{ab}$~\footnote{Email: kroll@physik.uni-wuppertal.de} and 
A.\ Sch\"afer$^{a}$~\footnote{Email: andreas.schaefer@physik.uniregensburg.de} \\[1em]
{\small \it a: Institut f\"ur Theoretische Physik, Universit\"at Regensburg, D-93040, Germany}
{\small {\it b: Fachbereich Physik, Universit\"at Wuppertal, D-42097 Wuppertal,
Germany}}\\
\end{center}
\vskip 5mm

\begin{abstract}
We analyze the time-like processes $\gamma\gamma\to\BbB$ and $p\bar{p}\to\gamma M$ at large
Mandelstam variables within the handbag approach for which the process amplitudes factorize
in hard partonic subprocesses and annihilation form factor. The latter represent moments
of baryon-antibaryon generalized parton distributions (GPDs). Symmetry relations restrict 
the number of independent annihilation form factors for the ground state baryons drastically. 
We determine these form factors from the present BELLE data on $\gamma\gamma\to\BbB$
with the help of simplifying assumptions. The knowledge of these form factors allow for 
predictions of $p\bar{p}\to \gamma M$ for various mesons which may be probed at FAIR. 
\end{abstract}

\section{Introduction}
For nearly twenty years hard exclusive reactions attracted much attention from both
Theoreticians and Experimentalists. This in particular true for the deeply 
virtual processes, Compton scattering and meson leptoproduction, where factorization
theorems tell us that, in the generalized Bjorken regime of large photon virtuality
and large energy, a process amplitude can be expressed as a convolution of hard, 
perturbatively calculable partonic subprocess amplitudes and soft hadronic matrix 
elements, parametrized as GPDs. In the course of 
time the accumulated data revealed the consistency of the theoretical concept, the 
so-called handbag approach, even though there are phenomena in the experimental data 
which seem to indicate the existence of strong corrections to the asymptotic 
factorization formula. It has been argued \ci{mueller-meskaukas} that these phenomena 
can, fully or partially, be accounted for by particular choices of the GPDs, i.e.\ 
they can be described by logarithms of the photon virtuality instead of powers as 
advocated for in \ci{GK3}. Whether or not this is possible for most or all these 
phenomena is yet unsolved. Deeply virtual exclusive processes have also been studied 
in the time-like region, e.g.\ $\gamma^*\gamma\to \pi^+\pi^-$ \ci{diehl-pire} or 
$p\bar{p}\to \gamma^*\pi$ \ci{pire}. Lack of data prevents the verification of this 
factorization approach till now.

Complementary to the deeply virtual reaction are the exclusive wide-angle processes
as for instance real Compton scattering \ci{rad,DFJK1} or their time-like counter part, 
the annihilation of two real photons into a pair of hadrons \ci{DKV2}-\ci{DK1}. 
The hard scale which permits factorization into hard and soft physics in this case, is 
provided by large Mandelstam variables $s\sim -t \sim -u$. The amplitudes are then 
expressed as a product of hard subprocess, $\gamma q\to \gamma q$ or 
$\gamma\gamma\to q\bar{q}$, and form factors representing moments of GPDs. In contrast 
to the deeply virtual reactions where factorization has rigorously been proven, 
factorization of the wide-angle reactions can only be shown to hold with the help of 
assumptions. Thus, for two-photon annihilation into a baryon-antibaryon pair 
factorization of the process amplitudes is achieved with the plausible assumption 
\ci{DKV3} that the process is dominated by configurations where the quark created in 
the subprocess, approximately moves in the direction of the baryon, whereas the 
antiquark moves in the direction of the antibaryon. This assumption is equivalent to 
the familiar valence quark approximation. By time reversal the amplitudes for 
two-photon annihilation into a proton-antiproton pair are the same as those for 
$p\bar{p}\to\gamma\gamma$ (up to eventual signs). As shown in \ci{kroll-schafer} the 
latter process can be generalized by replacing one of the photons in the final state 
by a meson.  

In this work we provide an update of \ci{DKV3} taking into account the new BELLE
data for $\gamma\gamma\to B\bar{B}$ \ci{kuo-05,abe} (where $B$ is a ground-state baryon). 
We now include the annihilation form factor for non-valence quarks in the numerical 
analysis as turned out to be necessary for 
$\gamma\gamma$ annihilations into a pair of pseudoscalar mesons \ci{DK1}. In contrast to
our previous work \ci{kroll-schafer} we are now going to explore in some detail the  
process $p\bar{p}\to \gamma V_L$ where $V_L$ is a longitudinally polarized vector meson. Both 
the classes of reactions, $\gamma\gamma\to B\bar{B}$ and $p\bar{p}\to \gamma V_L$, are linked 
to each other in so far as the same soft $B\bar{B}$ annihilation form factors occur. We think 
of our study as being timely since, in a few years from now, BELLE II and FAIR will be put 
into operation. In the first experiment the $\gamma\gamma\to B\bar{B}$ processes can be 
measured with high accuracy while at FAIR the proton-antiproton initiated reactions can be 
studied experimentally.
  
Plan of the paper is the following: In Sect.\ \ref{sec:handbag} we sketch the handbag approach
to wide-angle time-like exclusive processes. The symmetry relations among the annihilation 
form factors and the phenomenological determination of the set of independent form factors
are discussed in Sect.\ \ref{sec:form-factors}. The next section is devoted to
applications of the handbag approach to proton-antiproton annihilation into a photon and a 
meson and Sect.\ \ref{sec:J-Psi} to the special case of the $J/\Psi$. The paper ends with a 
summary and an outlook (Sect.\ \ref{sec:summary}).

\section{The handbag approach}
\label{sec:handbag}
In \ci{DKV3,freund} it has been argued that for large Mandelstam variables $s, -t$ and $-u$, 
the amplitudes for two-photon annihilations into pairs of ground state baryons factorize in 
a hard subprocess, $\gamma\gamma\to q\bar{q}$, and in soft annihilation form factors, 
$R_i^\gamma(s)$. In this so-called handbag approach the differential cross section takes the 
form
\be
\frac{d\sigma}{dt}(\gamma\gamma\to\BbB)\= \frac{4\pi\ale^2}{s^2}\,
                         \frac1{\sin^2{\theta}}
           \Big[ |R_V^\gamma(\BbB,s)|^2\,\cos^2{\theta}\,+\,
                       |R^\gamma_{\rm eff}(\BbB,s)|^2\Big]
\label{eq:gg-cross-section}
\ee
where $\theta$ is the scattering angle in the center-of-mass system and the effective 
annihilation form factor is short for the combination ($m_B$ is the mass of the baryon)
\be
|R^\gamma_{\rm eff}(s)|\=\sqrt{|R_A^\gamma(s)+R_P^\gamma(s)|^2+
                                    \frac{s}{4m_B^2}|R_P^\gamma(s)|^2}\,.
\label{eq:Reff}
\ee
The form factors represent the lowest moments of $\BbB$ distribution
amplitudes, $\Phi_i^q$, 
\be
F_{\BbB\,i}^q(s) \= \int_0^1 dz \Phi_{\BbB\,i}^q(z,\zeta,s)\,,
\ee
multiplied with the correspondent quark charges and summed over flavors 
($q=u, d, s$, $i=V,A,P$),
\be
R_i^\gamma(\BbB,s)\=e_u^2 F_{\BbB\,i}^u(s) + e_d^2 F_{\BbB\,i}^d(s) 
                        + e_s^2 F_{\BbB\,i}^s(s)\,.
\ee
Here $e_q$ is the charge of the quark of flavor $q$ in units of the positron
charge $e_0$ and~\footnote
{The exact definition of the $B\ov{B}$ distribution amplitudes, the time-like
 versions of generalized parton distributions, can be found in \ci{DKV3}. The 
scale dependence of the distribution amplitudes is not displayed for convenience.}  
The $\BbB$ distribution amplitudes and hence their moments are universal, 
i.e.\ process independent while the annihilation form factors are 
process-dependent flavor combinations of the moments $F^q_{\BbB\,i}$.

In the time-like region the skewness is defined by
\be
\zeta\=\frac{p^+}{p^+ + p^{\prime +}}
\ee
where $p^+$ and $p^{\prime +}$ are the light-cone plus components of the baryon 
and antibaryon momenta, respectively. The scattering-angle dependence in 
\req{eq:gg-cross-section} comes from the hard subprocess which is computed 
to lowest order of QED. By time-reversal invariance \req{eq:gg-cross-section} 
also holds for the process $\BbB\to \gamma\gamma$. 

As discussed in detail in \ci{DKV2,DKV3} the factorized cross section
\req{eq:gg-cross-section} is derived in a center-of-mass frame where the process 
takes place in the 1-3 plane and the outgoing hadrons move along the positive or 
negative 1-direction. In this frame both the hadron momenta have the same plus 
components and, hence, skewness is 1/2. In principle there is also a scalar
distribution amplitude and a corresponding scalar form factor. However, in a
$\zeta=1/2$ frame the scalar form factor decouples \ci{DKV3}.

The $q\bar{q}\to\BbB$ transitions and therefore the annihilation form factors 
can only be soft if the quark and the antiquark have small virtualities and 
momenta that are approximately equal to those of the baryon and antibaryon 
($z\simeq 1/2$), respectively. Corrections to this approximations are of order 
$\Lambda^2/s$ where $\Lambda$ is a typical hadronic scale of order $1\,\gev$. Thus, 
the dynamics we consider can be viewed as a time-like version of the Feynman mechanism.

\begin{figure}
\begin{center}
\includegraphics[width=0.4\tw]{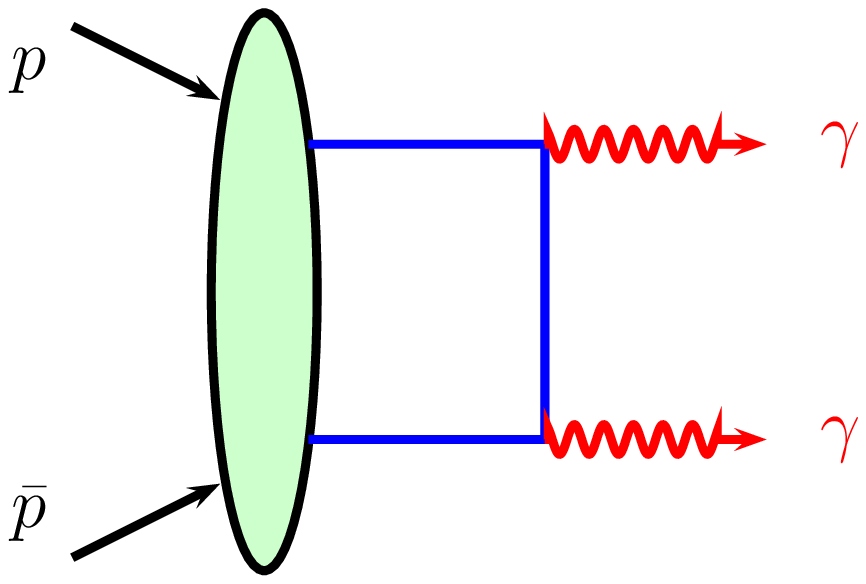}\hspace*{0.05\tw}
\includegraphics[width=0.4\tw]{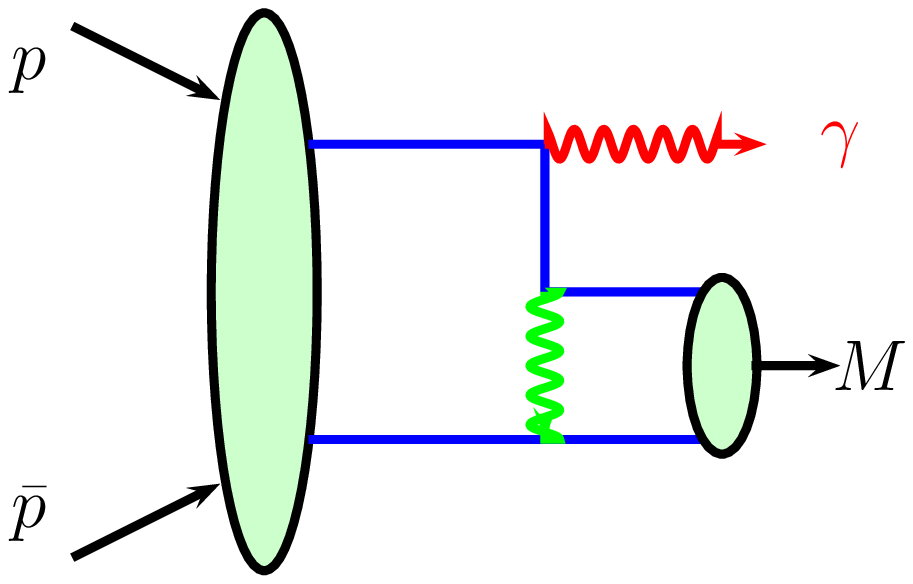}
\caption{\label{fig:graph} A sample Feynman graphs for $p\bar{p}\to\gamma\gamma$ 
and $p\bar{p}\to \gamma M$.}
\end{center}
\end{figure}

The handbag approach to $\gamma\gamma\to\BbB$ or better 
$\BbB \to\gamma\gamma$ can straightforwardly be generalized to processes 
for which one of the photons is replaced by pseudoscalar ($P$) or a longitudinally 
polarized vector meson ($V_L$) \ci{kroll-schafer}. Of experimental interest are
the processes with ingoing proton and antiproton
\ba
\frac{d\sigma}{dt}(p\bar{p}\to\gamma P) &=& \frac{\ale}{2s^3}\,
       \frac{|\bar{a}_P|^2}{\sin^4{\theta}}
       \Big[|R^P_{\rm eff}(s)|^2 + \cos^2{\theta}|R^P_V(s)|^2\Big]\,, \nn\\
\frac{d\sigma}{dt}(p\bar{p}\to\gamma V_L) &=& \frac{\ale}{2s^3}\,
       \frac{|\bar{a}_V|^2}{\sin^4{\theta}}
       \Big[\cos^2{\theta} |R^{V_L}_{\rm eff}(s)|^2 + |R^{V_L}_V(s)|^2\Big]\,.  
\label{eq:meson-cross-section} 
\ea
Since only the proton-antiproton initial state is experimentally feasible we
omit the label $p\bar{p}$ of the form factors for the ease of reading.
The form factor $R^M_{\rm eff}$ denotes the combination of $R_A^M$ and $R_P^M$
analogously to \req{eq:Reff}. The hard subprocess $q\bar{q}\to \gamma M$ is to 
be calculated perturbatively from the one-gluon exchange contribution of which
a typical Feynman graph is shown in Fig.\ \ref{fig:graph}. In this dynamical 
mechanism the meson is generated from its valence Fock component in collinear 
approximation. Consequently, the meson selects its valence (anti)quarks
from the (anti)proton. In terms of the universal moments of the generalized 
distribution amplitudes the annihilation form factors read 
\ba
R_i^{\,\rho^0, \pi^0}&=&\frac1{\sqrt{2}}\big(e_uF_i^u-e_dF_i^d)\,, \nn\\
R_i^{\,\omega_q, \eta_q}&=&\frac1{\sqrt{2}}\big(e_uF_i^u+e_dF_i^d)\,, \nn\\
R_i^{\,\omega_s,\eta_s}&=&e_sF_i^s\,.
\label{eq:gamma-M}
\ea
The $\eta_q (\omega_q)$ and $\eta_s (\omega_s)$ are the non-strange 
($(u\ub+d\db)/\sqrt{2}$) and strange ($s\sb$) components of the physical 
$\eta (\omega)$ and $\eta^\prime (\phi)$ mesons~\ci{FKS1} which are the basis states
in the quark-flavor basis. The mixing angle, $\Phi_P$, of the pseudoscalars in 
this basis amounts to $39.3^\circ$ \ci{FKS1,FKS2}. For the $\omega$ and $\phi$ system 
the corresponding mixing angle, $\Phi_V$, is known to be very small: from the 
$\phi\to \pi^0\gamma$ and $\omega\to\pi^0\gamma$ branching ratios \ci{PDG} one finds
a value of $3.3^\circ$ for the vector-meson mixing angle \ci{mixing}. 
For the vector-meson channels there is in principle also a contribution from the
gluonic subprocess $gg\to\gamma V_L$ along with the gluonic form factors
$F_i^g(s)$. However, this contribution is zero as has been note earlier in \ci{huang} 
for the space-like process $\gamma p\to V_Lp$ and in \ci{man-vat} for 
$\gamma g\to J/\psi g$ in the limit $M_{J/\psi}\to 0$.  

Evaluating the function $\bar{a}_M$ to leading order of perturbation 
theory, one finds 
\be
\bar{a}_M^{\rm coll} \= 4\pi\als(\mu_R) \frac{C_F}{N_c} f_M\langle
1/\tau\rangle_M
\label{eq:collinear}
\ee
in collinear approximation where $f_M$ is the meson's decay constant, 
$\langle 1/\tau\rangle_M$ is the $1/\tau$ moment of its distribution amplitude 
($\tau$ is the momentum fraction the quark entering the meson carries) and 
$\mu_R$ is an appropriate renormalization scale. The number of colors is denoted 
by $N_c$ and $C_F=(N_c^2-1)/(2N_c)$. The  $s-t$ crossed version of 
\req{eq:meson-cross-section} is the handbag result for wide-angle meson 
photoproduction as derived in \ci{huang}.

\section{Determination of the annihilation form factors}
\label{sec:form-factors}
The handbag mechanism only involves quark-antiquark intermediate states, 
$\gamma\gamma\to q\bar{q}\to \BbB$. Hence, isospin 2 (as well as $V$-spin 2)
transitions are absent. Similar to the case of $\gamma\gamma\to M\ov{M}$
\ci{DKV2,DK1} this dynamical selection rule in combination with
SU(3) flavor symmetry, in particular $U$-spin invariance, leads to relations
among the moments $F_{\BbB\,i}^q$. For each $i=V, A, P$  there are only three 
independent form factors \ci{DKV3} for which we choose
\be
F^q_i\=F^q_{p\bar{p} i}\,.
\ee
In terms of these independent moments or basic form factors the 
annihilation form factors for the $\gamma\gamma\to\BbB$ processes read
\ba
R_i^\gamma(\,p\bar{p}\,)&=& e_u^2 F_i^{\,u} + e_d^2 F_i^{\,d} + e_s^2 F_i^{\,s}\,,\nn\\
R_i^\gamma(\,n\bar{n}\,) &=& e_u^2 F_i^{\,d} + e_d^2 F_i^{\,u}+e_s^2 F_i^{\,s}\,,\nn\\ 
R_i^\gamma(\Sigma^+\ov{\Sigma}{}^-) &=& e_u^2 F_i^{\,u}+e_d^2 F_i^{\,s}
                                                     +e_s^2 F_i^{\,d}\,, \nn\\ 
R_i^\gamma(\Sigma^-\ov{\Sigma}{}^+) &=& e_u^2 F_i^{\,s} + e_d^2 F_i^{\,u} 
                                               + e_s^2 F_i^{\,d}\,, \nn\\ 
R_i^\gamma(\Sigma^0\ov{\Sigma}{}^0) &=& -\frac12(e_u^2+e_d^2)( F_i^{\,u} 
                 + F_i^{\,s})  - e_s^2 F_i^{\,d}\,,    \nn\\
R_i^\gamma(\,\Lambda\ov{\Lambda}\,) &=& 
            -\frac16(e_u^2+e_d^2) ( F_i^{\,u} + 4F_i^{\,d}+ F_i^{\,s})
            -\frac{e_s^2}3 (2F_i^{\,u} - F_i^{\,d} + 2F_i^{\,s})\,,   \nn\\
R_i^\gamma(\Lambda\ov{\Sigma}{}^0) &=& R_i^\gamma(\Sigma^0\ov{\Lambda}) =
               -\frac{\sqrt{3}}6(e_u^2-e_d^2)( F_i^{\,u} - 2F_i^{\,d}+ F_i^{\,s})\,, 
\label{eq:flavor-decomposition}
\ea                                            
up to corrections due to breaking of flavor symmetry which we ignore in this
work. The results for the form factors of the cascade particles for which flavor 
symmetry breaking is expected to be sizable, can be found in \ci{DKV3}. 

\begin{figure}
\begin{center}
\includegraphics[width=0.45\tw]{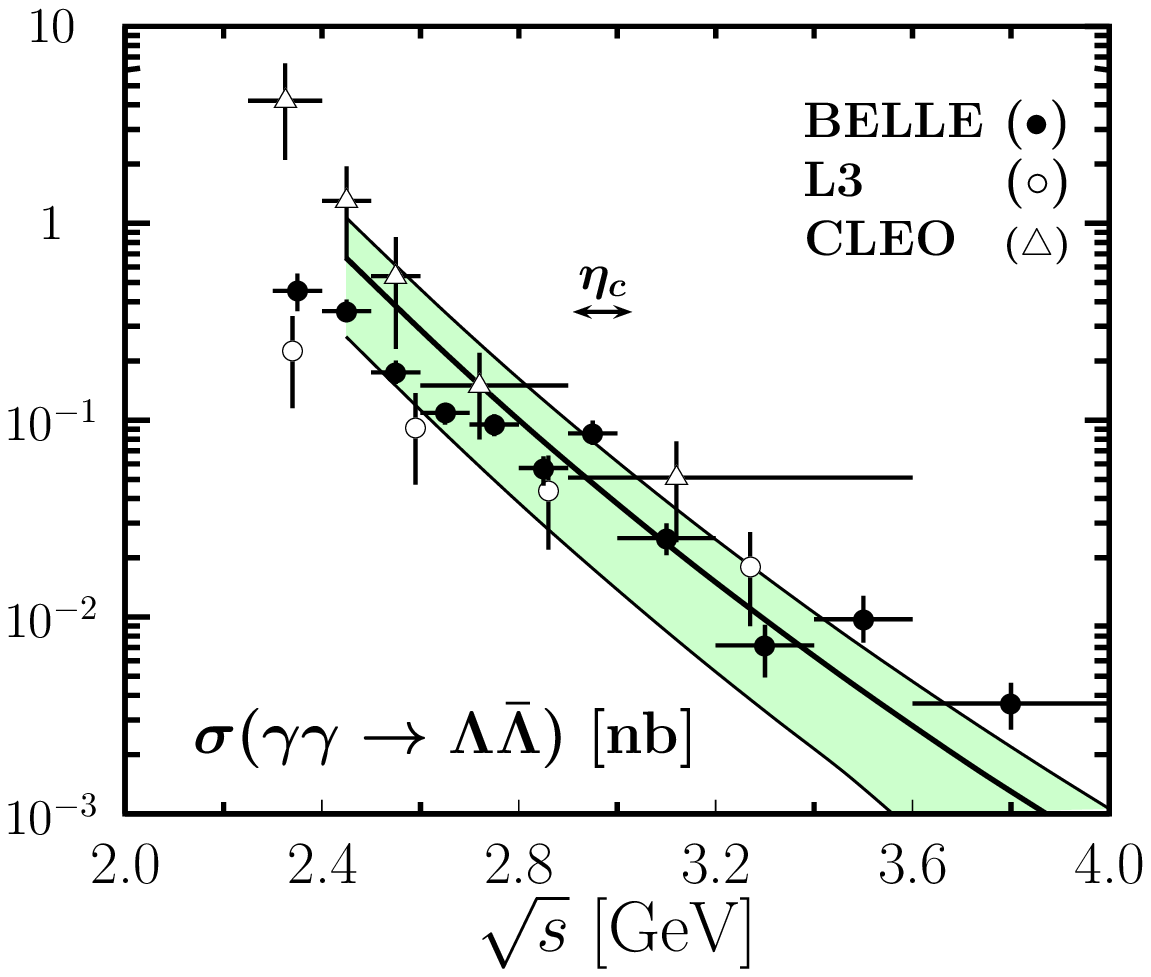}
\includegraphics[width=0.45\tw]{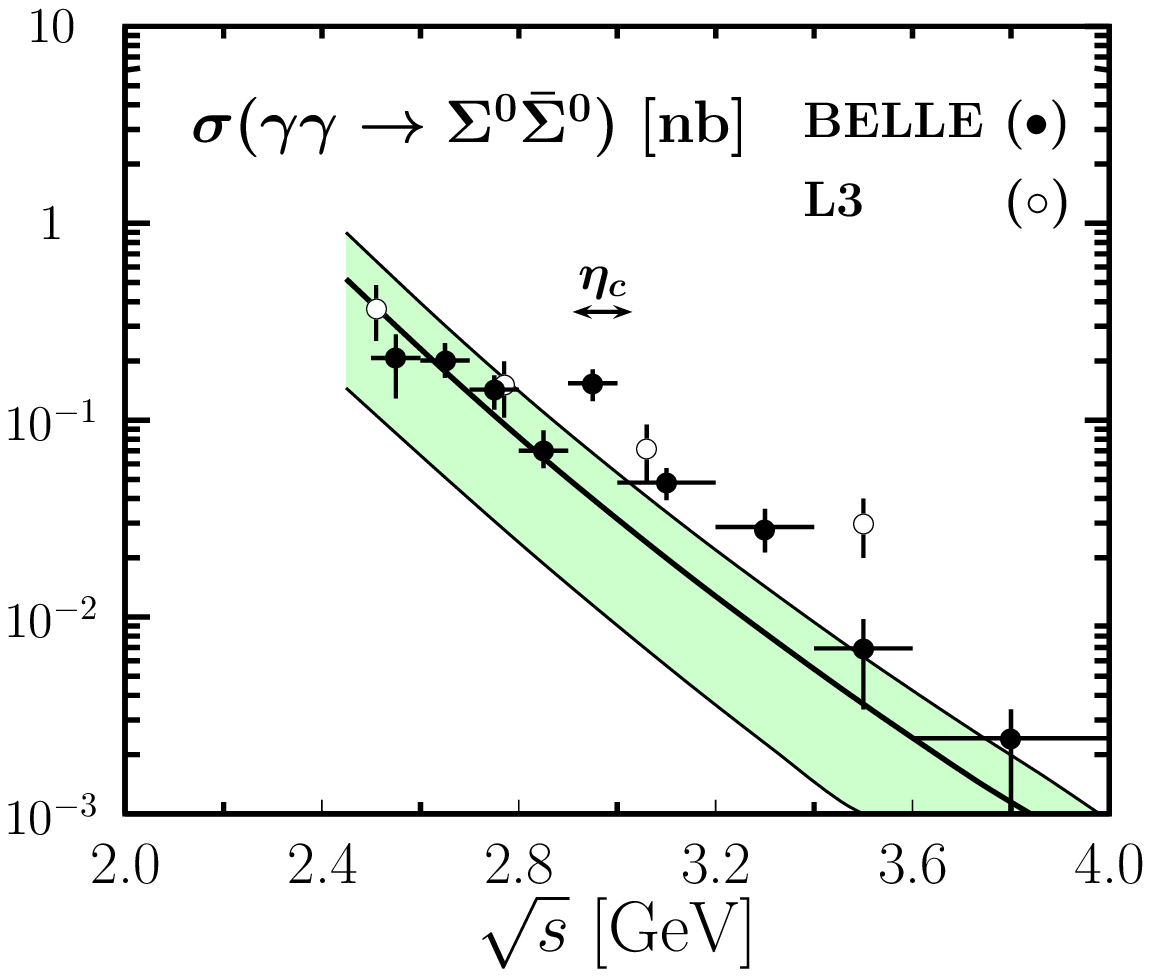}
\caption{\label{fig:hyperon} The integrated cross sections for
  $\gamma\gamma\to \Lambda\ov{\Lambda}$ (left) and $\gamma\gamma\to
  \Sigma^0\ov{\Sigma}{}^0$ (right) ($|\cos{\theta}|<0.6$). Data are taken from BELLE 
  \ci{abe} (preliminary), L3 \ci{L3} and CLEO \ci{CLEO}. The solid lines represent 
  our fit, evaluated from \req{eq:parameters}, and the shaded bands its uncertainties.} 
\end{center}
\end{figure}

In principle it is possible to extract the absolute magnitude of the moments, 
$F_i^{\,q}$, as well as their relative phases from a sufficiently large set of 
data on $\gamma\gamma\to\BbB$. From the angular dependence of the cross
sections one may separate the effective form factors from the vector ones
by a Rosenbluth type of separation. Measurements of the helicity correlation 
between the baryon and antibaryon would further allow to isolate $F^{\,q}_P$ 
from $F^{\,q}_A$. With regard to the accuracy of the present data on 
$\gamma\gamma\to\BbB$ this program is not feasible currently. In order to 
estimate the $F_i^{\,q}$ we therefore simplify by assuming
\be
F_i^{\,d}\=\rho_d\, F_i^{\,u}\,, \qquad F_i^{\,s}\=\rho_s\, F_i^{\,u}\,.
\label{eq:approx}
\ee
Due to the simplification of choosing $i$-independent factors $\rho_d$ and $\rho_s$
we can introduce a combination of moments $F_i^q$ analogously to \req{eq:Reff} which 
is related to $R^\gamma_{\rm eff}$ by
\be
F^u_{\rm eff}\= R^\gamma_{\rm eff}(p\bar{p})
                     \big[e_u^2+e^2_d\rho_d+e_s^2\rho_s\big]^{-1}\,.
\ee
We fit the basic form factors, i.e.\ $R^\gamma_{\rm eff}$, $|R_V^\gamma|$, $\rho_d$ and $\rho_s$,
to the available data. Relative phases between the basic form factor are ignored
since the available data do not allow to fix them. The accurate BELLE data on the 
$\gamma\gamma\to p\bar{p}$ differential and integrated cross sections \ci{kuo-05} 
determine  $R^\gamma_{\rm eff}(p\bar{p})$ and $R_V^\gamma(p\bar{p})$ 
while the cross section data, integrated over the wide-angle region
($|\cos{\theta}|\leq 0.6$),  for the processes $\gamma\gamma\to\Lambda\ov{\Lambda},\,
\Sigma^0\ov{\Sigma}{}^0$ from BELLE \ci{abe}, L3 \ci{L3} and CLEO \ci{CLEO}
fix in addition $\rho_d$ and $\rho_s$. Data are taken into account only for $s$ 
larger than $7\,\gev^2$ (and leaving out the region of the $\eta_c$ formation). For 
$s$ smaller than about $7\,\gev^2$ the $\gamma\gamma\to p\bar{p}$ differential cross 
section reveals a maximum at a scattering angle of $90^\circ$ \ci{kuo-05} which is 
in conflict with the properties of the handbag dynamics. A best fit provides 
($s_0=10.4\,\gev^2$)
\ba
s^2R^\gamma_{\rm eff} &=& (3.12\pm 0.3) \left(s/s_0\right)^{-1.10\pm 0.10} \gev^4\,, \nn\\
s^2|R^\gamma_V| &=& (8.82\pm 0.8)  \left(s/s_0\right)^{-1.10\pm 0.10} \gev^4\,, \nn\\
\rho_d &=& 0.55\pm 0.15\,,  \nn\\ 
\rho_s&=&-(0.13 \pm 0.11)s_0/s\,.
\label{eq:parameters}
\ea
Admittedly the fit is to be regarded as a rough estimate, the present data do not 
allow for a more accurate determination of the annihilation form factors. The fit 
to the $\gamma\gamma\to p\bar{p}$ differential and integrated cross section is 
practically the same as our previous fit presented in \ci{kroll-schafer}. We 
therefore refrain from showing plots of the $\gamma\gamma\to p\bar{p}$ observables 
here. The fit to the $\Lambda\ov{\Lambda}$ and $\Sigma^0\ov{\Sigma}{}^0$ cross 
sections is compared to the data in Fig.\ \ref{fig:hyperon}. Note the sign of 
$\rho_s$ which is the same as the relative sign between the valence and non-valence 
form factors in the case of two-photon annihilations into pairs of mesons \ci{DK1}. 

The cross sections for other $\BbB$ channels can easily be worked out
from \req{eq:flavor-decomposition} and \req{eq:approx}. Their ratios to the $p\bar{p}$
cross section is fixed by quark charges and the parameters $\rho_d$ 
and $\rho_s$. Thus for instance, the ratio of the $n\bar{n}$ and $p\bar{p}$ cross sections
is 0.49 at $s\simeq 30\,\gev^2$ and the $\Sigma^-\ov{\Sigma}^+/p\bar{p}$ ratio is 0.09.
The cross section for $\gamma\gamma\to\Lambda\ov{\Sigma}{}^0$ is very small, suppressed
by about a factor 1000 as compared to the $p\bar{p}$ one. The non-valence
form factor is very important in this case leading to a stronger energy dependence of
that cross section than the other ones exhibit.  

Before closing this section a remark is in order concerning an alternative mechanism
for two-photon annihilation into a baryon-antibaryon pair. This is the so-called
perturbative QCD factorization scheme \ci{BL,CZ} which also holds for large
Mandelstam variables. In this factorization scheme the valence Fock components
of the proton and the antiproton are fully resolved in three quarks and antiquarks,
respectively. Higher Fock components are suppressed by powers of the hard scale $\sim s$.
In contrast to the handbag mechanism where there is only one active quark and antiquark,
all quarks and antiquarks of the valence Fock components participate in the hard 
subprocess which necessitates the exchange of two hard gluons to lowest order of 
perturbative QCD. The baryons are described by ordinary three-quark distribution 
amplitudes, i.e.\ by wave functions integrated over quark transverse momenta. It 
turns out, however, that this factorization scheme fails badly with the normalization 
of the cross section \ci{farrar}, it is way below experiment, in particular if 
distribution amplitudes are used that are close to the asymptotic form 
which are favored by phenomenology \ci{bolz}, lattice QCD \ci{braun} and by light-cone
sum rules \ci{braun13}~\footnote
{There is a variant of the perturbative QCD mechanism in which a baryon is viewed as 
being composed of a quark and a diquark \ci{schweiger}. This variant leads
to much better agreement with experiment.}. 
Despite this apparent failure the perturbative QCD factorization scheme is presumably 
the correct mechanism for $s\sim -t\sim -u \to\infty$ while the handbag contribution 
dominates the process $\gamma\gamma\to B\ov{B}$ for large but not asymptotically large 
Mandelstam variables~\footnote
{As pointed out in \ci{close} quark-hadron duality provides additional justification for the 
handbag approach in this kinematic domain - contributions from cat's ears topology
where the two photons couple to different quarks, are likely suppressed.}.
The perturbative QCD factorization scheme predicts an energy 
dependence of the integrated $B\ov{B}$ cross sections as $s^{-5}$ while, experimentally, 
the $p\bar{p}$ cross sections falls as  $\simeq s^{-7.2}$. The energy dependence of the 
phenomenological annihilation form factors take that experimental result into account. 
Thus, our analysis is consistent with the dominance of the perturbative QCD factorization 
scheme at very large energies.

\section{\boldmath  $p\bar{p}\to \gamma M$ \unboldmath \bf phenomenology} 
\label{sec:meson}
As reported in \ci{kroll-schafer} the handbag approach works quite well for the 
$\gamma\pi^0$ channel as far as the energy and scattering-angle dependence is concerned. 
While the leading-order, collinear result \req{eq:collinear} reproduces these 
features it fails with the normalization of the cross section by order of magnitude
as is also the case for the space-like process, photoproduction of the $\pi^0$ 
meson off protons \ci{huang}. It is therefore suggestive to assume that the 
handbag factorization holds as well for other photon-meson channels with subprocesses
evaluated to leading-order, collinear accuracy and a normalization that is sufficiently
enhanced by a more general mechanism. An example of such a mechanism is the insertion 
of an infinite number of fermionic loops in the hard gluon propagator \ci{belitsky}. 
Such a mechanism likely leads to an enhancement of the normalization independent on 
the produced meson. In any case a channel independent-enhancement of the normalization 
will be assumed throughout this paper. In \ci{kroll-schafer} the normalization
has been fitted to the Fermilab data on $p\bar{p}\to \gamma\pi^0$ \ci{e760}.
Lack of data prevents a similar procedure for the $V_L$ channels.
 
\begin{figure}
\begin{center}
\includegraphics[width=0.40\tw]{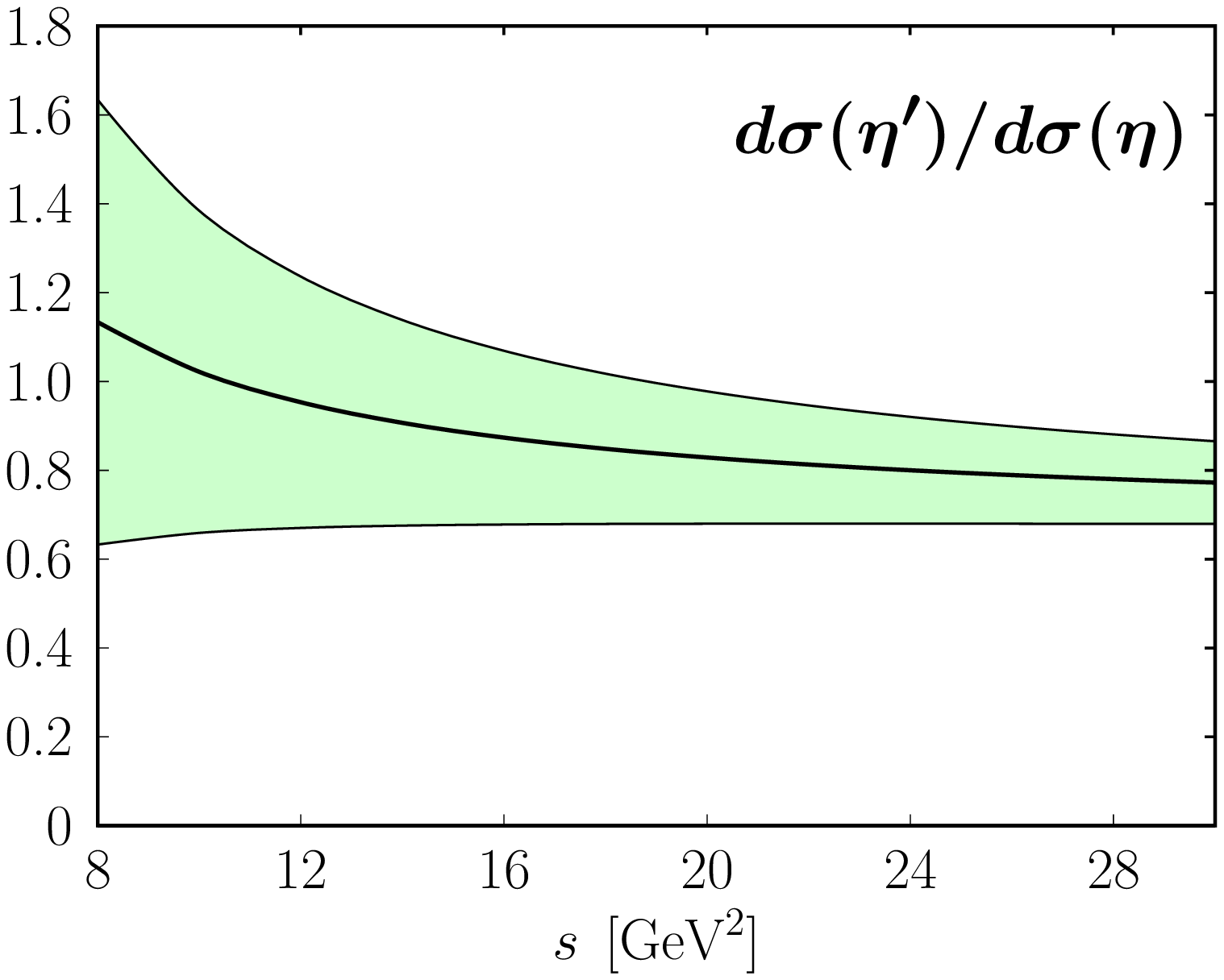}\hspace*{0.05\tw}
\includegraphics[width=0.40\tw]{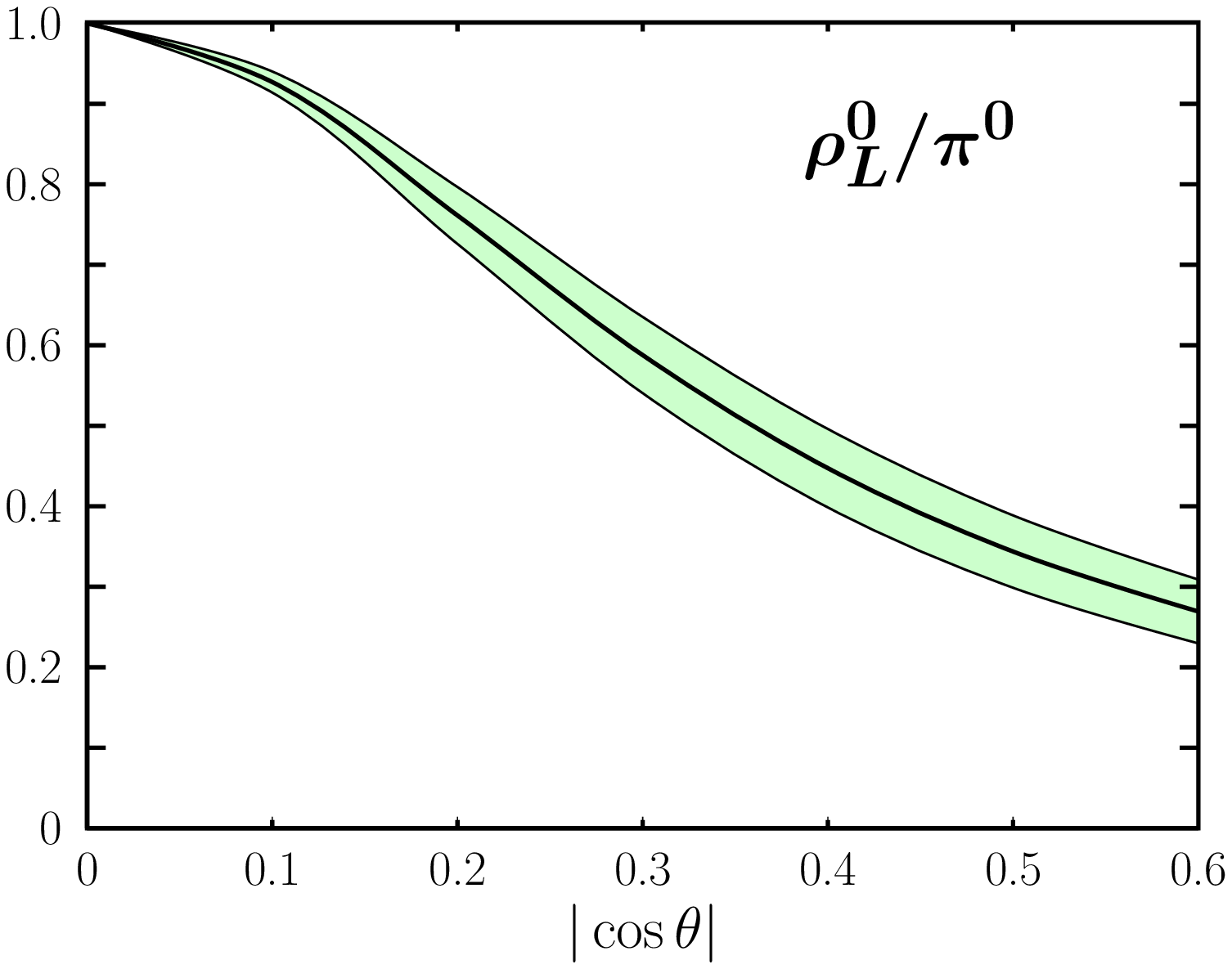}
\caption{\label{fig:rho-pi} Left: The ratio of the $\eta^\prime$ and $\eta$ cross 
sections versus $s$. Right: The ratio of the $p\bar{p}\to\gamma\rho^0_L$ and 
$p\bar{p}\to\gamma\pi^0$ cross sections, scaled by their $90^\circ$ values, versus
$|\cos{\theta}|$. For other notations refer to Fig.\ \ref{fig:hyperon}.}
\end{center}
\end{figure}

In \ci{kroll-schafer} the non-valence form factor has not been taken into account in 
the analysis. In this situation the $\eta$ and the $\eta^\prime$ meson are solely 
generated through the $\eta_q$ state. The inclusion of the non-valence form factor 
modifies the result for the ratio of the $\eta$ and $\eta^\prime$ cross sections presented 
in \ci{kroll-schafer}; the physical mesons can also be created by the $s\sb$ intermediate
state. Neglecting the two-gluon Fock components of the $\eta$ and $\eta^\prime$ mesons, we 
find for the $\eta^\prime - \eta$ cross section ratio 
\be
\frac{d\sigma/dt(p\bar{p}\to\gamma \eta^\prime)}{d\sigma/dt(p\bar{p}\to\gamma \eta)}
       \=\tan^2{\Phi_P}\left|\frac{1 + \kappa_P\cot{\Phi_P}}{1-\kappa_P\tan{\Phi_P}}\right|^2
\ee
where
\be
\kappa_P\=\sqrt{2}\, \frac{f_s\langle1/\tau\rangle_{\eta_s}}
             {f_q\langle1/\tau\rangle_{\eta_q}}\,\frac{e_s\rho_s}{e_u+e_d \rho_d}\,.
\ee
The ratio of the $\eta$ and $\eta^\prime$ cross section is independent on the scattering
angle. This is also case for the $\eta-\pi^0$ ratio as can easily be checked.
The decay constants, $f_q$ and $f_s$, for the basis states, $\eta_q$ and $\eta_s$, have 
been estimated in \ci{FKS1,FKS2}: $f_q=1.07 f_\pi$ and $f_s=1.34 f_\pi$. The corresponding
distribution amplitudes do not differ much otherwise large OZI rule violations would occur
in conflict with experiment \ci{passek}. Taking all this information into account,
we estimate $\kappa_P$ to amount to $\simeq 0.16 s_0/s$. The $\etp - \eta$ cross section 
ratio is shown in Fig.\ \ref{fig:rho-pi}. For $s\to\infty$ it is simply given by 
$\tan^2{\Phi_P}$. This is the result obtained in \ci{kroll-schafer}. An analogous result 
is obtained for the $\eta/\pi^0$ ratio:
\be
\frac{d\sigma/dt(p\bar{p}\to\gamma \eta)}{d\sigma/dt(p\bar{p}\to\gamma \pi^0)}
       \=\cos^2{\Phi_P}\Big[\frac{f_q\langle 1/\tau\rangle_{\eta_q}}{f_\pi\langle 1/\tau\rangle_\pi}
         \,\frac{e_u+e_d\rho_d}{e_u-e_d\rho_d}\Big]^2
            \left|{1-\kappa_P\tan{\Phi_P}}\right|^2\,.
\label{eq:eta-pi0}
\ee
This ratio is clearly smaller than 1. Assuming equal distribution amplitudes for the 
$\eta_q$ and the pion, we obtain a value of 0.17 for the $\eta/\pi^0$ ratio at $s$ about 
$10\,\gev^2$. This value increases slowly with $s$ up to a value of 0.22. Such small values 
of the $\eta/\pi^0$ ratio are also expected for deeply virtual kinematics \ci{GK6} and 
have been observed experimentally \ci{clas-eta-pi0}.

The cross sections for the $\gamma V_L$ channels differ from that for the $\gamma\pi^0$
channel by the fact that $R^{V_L}_{\rm eff}$ now goes along with the $\cos{\theta}$
dependence instead of $R^{V_L}_V$. As an example we show  in Fig.\ \ref{fig:rho-pi}
the angle dependence of the ratio of the $\gamma\rho_L^0$ and $\gamma\pi^0$ cross 
sections, scaled by the cross sections at $\theta=90^\circ$. The scaled ratio reads
\be
\frac{d\sigma/dt(p\bar{p}\to\gamma \rho^0_L)}{d\sigma/dt(p\bar{p}\to\gamma \pi^0)}\Big/
\frac{d\sigma/dt(p\bar{p}\to\gamma \rho^0_L,90^\circ)}
           {d\sigma/dt(p\bar{p}\to\gamma \pi^0,90^\circ)}
\= \frac{1 + \cos^2{\theta}(F^u_{\rm eff}/|F^u_V|)^2}
          {1 + \cos^2{\theta}(|F^u_V|/F^u_{\rm eff})^2}\,.
\ee
It does not depend on energy.

Since the $\omega - \phi$ mixing angle is very small and the valence form factors are 
substantially larger than the non-valence one the strange quark admixture to the $\omega$
mesons can safely be ignored. Therefore, we find the simple result for the 
$\omega/\rho^0$ ratio
\be
\frac{d\sigma/dt(p\bar{p}\to\gamma \omega_L)}{d\sigma/dt(p\bar{p}\to\gamma \rho_L^0)}
\simeq \left(\frac{f_\omega\langle 1/\tau\rangle_\omega}
               {f_\rho\langle 1/\tau\rangle_\rho}\right)^2\,
              \left|\frac{e_u+e_d \rho_d}{e_u-e_d \rho_d}\right|^2\,.
\ee
This is the analogue of \req{eq:eta-pi0} for a zero mixing angle. 
For the $\phi$ channel, on the other hand, we obtain~\footnote{
Since the mixing angle is so small we approximate the decay constants of
the non-strange and strange Fock components of the $\omega$ and $\phi$ mesons by 
the phenomenological values of the physical mesons.} 
\be
\frac{d\sigma/dt(p\bar{p}\to\gamma \phi_L)}{d\sigma/dt(p\bar{p}\to\gamma \omega_L)}
= \tan^2{\Phi_V}\left| 1+ \kappa_V \cot{\Phi_V}\right|^2
\ee
where 
\be
\kappa_V\=\sqrt{2}\, \frac{f_\phi\langle1/\tau\rangle_{\phi}}
             {f_\omega\langle1/\tau\rangle_{\omega}}\,\frac{e_s\rho_s}{e_u+e_d \rho_d}\,.
\ee

\begin{figure}
\begin{center}
\includegraphics[width=0.40\tw]{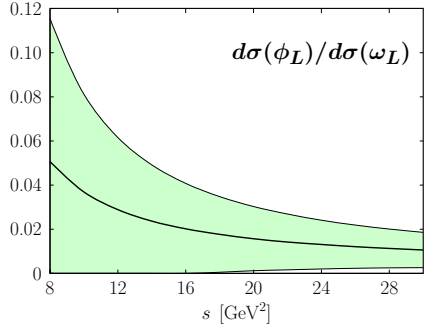}
\caption{\label{fig:mixing} The ratio of the $\phi_L$ and $\omega_L$ cross sections 
versus $s$. For other notations refer to Fig.\ \ref{fig:hyperon}.}
\end{center}
\end{figure}

For $s\simeq s_0$ the strange quark content of the
proton, embedded in the non-valence form factor $F_i^s$, is strong enough to generate
the $\phi$ meson through its $s\sb$ Fock component. For large $s$, on the other hand,
the non-valence form factor is suppressed and the $\phi$ is generated by the non-strange
quarks through mixing which results in
\be
\frac{d\sigma/dt(p\bar{p}\to\gamma \phi_L)}{d\sigma/dt(p\bar{p}\to\gamma \omega_L)}
       \stackrel{s\to\infty}{\longrightarrow} \tan^2{\Phi_V}\,.
\ee

For an quantitative estimate of the vector meson cross sections we take $f_{\rho^0}=209\,\mev$, 
$f_\omega=187\,\mev$ and $f_\phi=221\,\mev$ \ci{beneke-neubert} and the Gegenbauer coefficients 
of the $\rho^0$ and $\phi$ mesons from \ci{ball-braun} ($a_2^\rho=0.2$, $a_2^\phi=0$ at the 
scale of $\mu_0=1\,\gev$). For the $\omega$ meson the distribution amplitude is unknown and 
we assume that it equals that of the $\rho$ meson. The scale dependent $1/\tau$ moment of a 
distribution amplitude is given by
\be
\langle 1/\tau\rangle\=3 \Big[1+a_2\big(\als(\mu_R)/\als(\mu_0)\big)^{50/81}\Big]\,.
\ee
For the renormalization (and factorization) scale we take the average gluon virtuality
in the subprocess $\mu_R=\sqrt{s}/2$ ($\Lambda_{\rm QCD}=181\,\mev$). Using these parameters, 
we estimate the $\omega/\rho^0$ ratio to amount to about $0.26\pm 0.09$ independent on energy.
The $\phi/\omega$ cross section ratio is displayed in Fig.\ \ref{fig:mixing}.

The generalization of our approach to transversely polarized vector mesons is 
principally possible. In this case one has to consider the subprocess 
$q\bar{q}\to\gamma V_T$ with equal and opposite quark and antiquark helicities. In the 
first case this requires the introduction of new $p\bar{p}$ distributions (time-like 
versions of the transversity GPDs), and, hence, a new set of associated annihilation form 
factors. For opposite helicities higher-twist meson distributions are required. The analysis 
of $p\bar{p}\to\gamma V_T$ is beyond the scope of this work.

\section{Remarks on \boldmath $J/\psi$\unboldmath production}
\label{sec:J-Psi}
One may also consider the process $p\bar{p}\to\gamma J/\psi$. The quark contribution to 
this process, i.e.\ the subprocess $c\bar{c}\to \gamma J/\psi$, is to be calculated 
analogously to the above discussed light-quark initiated processes except that the charm
quark mass, $m_c$, is to be taken into account. The annihilation form factors $F_i^c$ are 
expected to be very small and to fall off with energy even more rapidly than $F_i^s$. 
Thus this contribution is likely very small as is the case for the space-like process, 
photoproduction of the $J/\psi$. As for the latter process (see e.g.\ \ci{ivanov}) 
there is also the possibility of the emission of a pair of gluons from the proton and 
antiproton which goes along with gluonic $p\bar{p}$ distribution amplitudes and  
associated gluonic annihilation form factors, $F_i^g(s)$.
The structure of the $J/\psi$ cross section is similar to \req{eq:meson-cross-section}
with however different perturbative coefficients multiplying the gluonic annihilation
form factors. In addition there is a contribution from configurations for which the
gluons have the same helicity instead as opposite helicity. This contribution goes 
along with with a form factor being related to a time-like gluonic transversity GPD.

In principle the gluonic annihilation form factors can be determined from data on the
$p\bar{p}\to\gamma J/\psi$ cross section in full analogy to the analysis of 
$\gamma\gamma \leftrightarrow p\bar{p}$. With the gluonic form factors at hand it is 
possible to estimate the cross sections for reactions like $p\bar{p}\to\gamma \Upsilon$ 
or $p\bar{p}\to\gamma \Psi^\prime$. At FAIR such reactions can be measured.  
Finally we note that for such measurements large $-t$ and $-u$ are not required since 
the mass of the heavy quark provides a hard scale.

\section{Summary and Outlook} 
\label{sec:summary}

We have investigated two classes of reactions, $\gamma\gamma\to\BbB$ and 
$p\bar{p}\to\gamma M$, for wide-angle kinematics, i.e.\ at large Mandelstam variables
$s, -t$ and $-u$. Within the handbag mechanism these reactions are complementary  
in so far as the soft physics information required for their description within a 
factorization approach, are encoded in the same set of annihilation form factors which 
represent lowest moments of $\BbB$ distribution amplitudes, time-like versions of GPDs. 
In contrast to previous studies of these processes \ci{DKV3,kroll-schafer} a 
non-valence form factor is taken into account in the numerical analysis. From the 
present $\gamma\gamma\to\BbB$ 
data \ci{kuo-05,abe} the annihilation form factors can be determined with the help of 
a few simplifying assumptions. The uncertainties of the form factors are however
rather large. More and better data are required for an improvement. Using these
form factors we are in the position to give a number of predictions for various
$p\bar{p}\to\gamma M$ channels, in particular for longitudinally polarized vector
mesons. These predictions can be probed at FAIR. This may lead to a better understanding
of the wide-angle exclusive processes and to a more precise knowledge of the 
annihilation form factors, in particular to the role of strangeness in the proton 
for $z$ close to $1/2$ in the time-like region which corresponds to large $x$ in the 
space-like region. This seems to be important in view of the recent controversy about 
the strangeness content of the nucleon. Traditional PDF-fits assume $s_p(x)=\bar s_p(x)$ 
to be substantially smaller than the light quark sea distribution with approximately the 
same $x$ dependence, while recent ATLAS data \cite{atlas} suggest that there is no such 
suppression at small $x$. In addition earlier HERMES data \cite{hermes-strange} suggest 
that $s(x)$ is much smaller than usually assumed at large $x$. As the strange content of 
the proton enters $W^{\pm}$-production cross sections at LHC this controversy has to be 
settled. We conclude that it would be good if measurements of $p\bar{p}\to\gamma M$ 
at wide-angle kinematics had a rather high priority within the FAIR program.

\end{document}